\documentclass[a4paper,fleqn,usenatbib]{mnras}
\usepackage{newtxtext,newtxmath}
\usepackage[T1]{fontenc}
\usepackage{ae,aecompl}

\usepackage{graphicx}	% Including figure files
\usepackage{amsmath}	% Advanced maths commands
\usepackage{amssymb}	% Extra maths symbols
\usepackage{verbatim,graphicx,epsfig,dcolumn,xcolor}

\newcolumntype{.}{D{.}{.}{4}}
\newcolumntype{,}{D{.}{.}{2}}
\newcolumntype{;}{D{.}{.}{1}}

\newcommand{\appropto}{\mathrel{\vcenter{
  \offinterlineskip\halign{\hfil$##$\cr
    \propto\cr\noalign{\kern2pt}\sim\cr\noalign{\kern-2pt}}}}}

\title[CO around 47 Tuc V3]{Circumstellar CO in metal-poor stellar winds: the highly irradiated globular cluster star 47 Tucanae V3}

\author[I. McDonald et al.]{
I.~McDonald,$^{1}$\thanks{E-mail: iain.mcdonald-2@manchester.ac.uk}
M.L.~Boyer,$^{2}$
M.A.T.~Groenewegen,$^{3}$
E.~Lagadec,$^{4}$
A.M.S.~Richards,$^{1}$ \newauthor
G.C.~Sloan,$^{2,5}$ and
A.A.~Zijlstra$^{1,6}$
\\
$^{1}$Jodrell Bank Centre for Astrophysics, Alan Turing Building, Manchester, M13 9PL, UK\\
$^{2}$Space Telescope Science Institute, 3700 San Martin Dr., Baltimore, MD, 21218, USA\\
$^{3}$Koninklijke Sterrenwacht van Belgi\"e, Ringlaan 3, 1180, Brussels, Belgium\\
$^{4}$Laboratoire Lagrange, Universit\'e C\^ote d'Azur, Observatoire de la C\^ote d'Azur, CNRS, boulevard de l'Observatoire, CS 34229, 06304 Nice Cedex 4, France\\
$^{5}$University of North Carolina Chapel Hill, Chapel Hill NC 20599-3255\\
$^{6}$Department of Physics \& Laboratory for Space Research, University of Hong Kong, Pokfulam Road, Hong Kong
}

\date{Accepted XXX. Received YYY; in original form ZZZ}

\pubyear{2015}

\begin{document}
\label{firstpage}
\pagerange{\pageref{firstpage}--\pageref{lastpage}}
\maketitle

\begin{abstract}
We report the first detection of circumstellar CO in a globular cluster.  Observations with ALMA have detected the CO $J$=3--2 and SiO $v$=1 $J$=8--7 transitions at 345 and 344 GHz, respectively, around V3 in 47 Tucanae (NGC 104; [Fe/H] = --0.72 dex), a star on the asymptotic giant branch. The CO line is detected at 7$\sigma$ at a rest velocity $v_{\rm LSR}$ = --40.6 km s$^{-1}$ and expansion velocity of 3.2 $\pm$ $\sim$0.4 km s$^{-1}$. The brighter, asymmetric SiO line may indicate a circumstellar maser. The stellar wind is slow compared to similar Galactic stars, but the dust opacity remains similar to Galactic comparisons. We suggest that the mass-loss rate is set by the levitation of material into the circumstellar environment by pulsations, but that the terminal wind-expansion velocity is determined by radiation pressure on the dust: a pulsation-enhanced dust-driven wind. We suggest the metal-poor nature of the star decreases the grain size, slowing the wind and increasing its density and opacity. Metallic alloys at high altitudes above the photosphere could also provide an opacity increase. The CO line is weaker than expected from Galactic AGB stars, but its strength confirms a model that includes CO dissociation by the strong interstellar radiation field present inside globular clusters.
\end{abstract}
%TOTAL: 197 words


%%%%%%%%%%%%%%%%%%%%%%%%%%%%%%%%%%%%%%%%%%%%%%%%%%

%%%%%%%%%%%%%%%%% BODY OF PAPER %%%%%%%%%%%%%%%%%%

\section{Introduction}

Stars on the asymptotic giant branch (AGB) are among the major sources of chemical enrichment in the Universe, alongside supernovae \citep[e.g.][]{KL14}. However, the mass loss that controls this enrichment is poorly understood \citep[e.g.][]{HO18}. AGB stars become unstable to pulsation, which shocks and levitates the outer atmosphere, allowing dust to form. Radiation pressure on these grains, and collisional coupling with the surrounding gas, forces a wind from the star. However, particularly in low-luminosity, oxygen-rich or metal-poor stars, the opacity needed to drive a wind cannot come from absorption and re-radiation of stellar light alone \citep{Woitke06b}. Scattering by large grains has been invoked as a solution \citep{Hoefner08,NTI+12}, but these are conceptually hard to grow around oxygen-rich, metal-poor stars in particular, due to the low fractional abundance of refractory elements. Mass loss in increasingly metal-poor environments is expected to become progressively limited to carbon stars \citep[e.g.][]{DCDAV+13}, unless additional, metallicity-independent mechanisms of mass loss are invoked (e.g., magnetic fields, active in red giant branch stars, may retain a role if radiation-driven winds fail; \citealt[cf.][]{DHA84}), or unless additional sources of opacity can be found.

Observations of metal-poor systems, however, indicate that oxygen-rich, metal-poor stars can be prodigious producers of dust \citep[e.g][]{BMB+15,BMG+17}, and there appears relatively little metallicity dependence in the integrated mass-loss rate from stars, under the limited range of masses, metallicities and evolutionary states that we can test \citep[e.g.][]{vLBM08,MZ15b}. However, an important observational unknown is the terminal expansion velocity ($v_\infty$) of metal-poor stellar winds. If winds remain dust driven, this velocity declines with metallicity \citep[e.g][]{vL00}; otherwise, additional sources of energy are needed to maintain the wind. CO, the most stable and abundant metal molecule, provides a good tracer for $v_\infty$. Circumstellar CO observations in the Magellanic Clouds have probed only the brightest, most massive stars on the AGB, which are the most metal-rich stars in an only mildly metal-poor environment \citep{GVM+16,MSS+16,GvLZ+17}. Their behaviour cannot be extrapolated to stars with lower luminosities or metallicities, where wind-driven outflows will be more difficult.

Globular clusters provide a laboratory for such objects. Dust has been spectroscopically confirmed around AGB stars down to at least [Fe/H] $\sim$ --1.6 dex \citep{BMvL+09,SMM+10,MvLS+11}, and claims exist within clusters of lower metallicity \citep{OFF+14}. The globular cluster 47 Tucanae (NGC 104) is particularly well studied. Though not metal-poor by conventional definition \citep{BC05}, 47 Tuc is more metal-poor ([Fe/H] = --0.72 dex; \citealt{Harris96}\footnote{The 2010 edition is used throughout.}) than young stars in the Magellanic Clouds. It is populous, and close ($\sim$4.5 kpc; \citealt{Harris96}). Its numerous dust-producing AGB stars \citep{LPH+06,vLMO+06,MBvLZ11} provide an important testbed for studying mass loss from low-luminosity, oxygen-rich, metal-poor stars.

Previous observations of AGB stars in 47 Tuc with the Atacama Large Millimeter Array (ALMA) could not detect the CO $J$=2-1 transition. \citet{MZL+15} argued that was due to rapid photo-dissociation of the CO envelope by a harsh interstellar radiation field (ISRF). \citet{MZ15a} showed this field could be strong enough to ionise any intra-cluster medium (ICM) and boil it off the cluster. In this case, higher transitions (CO $J$=3--2), excited closer to the star, should be more easily observed. We selected the third-brightest AGB star, V3, for re-observation, as it is the furthest from the cluster centre, hence least subject to ionising radiation. V3 (3540 K, 4590 L$_\odot$, log($g$) = --0.1 dex, $P$ = 221 days, $\delta V$ = 3.5 mag; \citealt{LNH+14}) exhibits an unusual dust spectrum, devoid of silicate features \citep{MBvLZ11}. Its mean radial velocity ($v_{\rm LSR} \approx -38$ km s$^{-1}$; \citealt{LW05})\footnote{The velocity is approximate because pulsations vary the optical velocity by $\pm\sim$10 km s$^{-1}$.} separates it from the cluster mean ($v_{\rm LSR}$ = --26.7 km s$^{-1}$; \citealt{Harris96}), allowing clear association with V3 as the source. The \citet{GaiaDR2} places its planar motion as radially away from the cluster centre at 10.7 $\pm$ 5.1 km s$^{-1}$, so its total space motion ($\sim$18 km s$^{-1}$) is marginally higher than the cluster central velocity dispersion ($\sim$13.5 km s$^{-1}$; inferred from \citealt{Harris96}).

\section{Observations \& Results}

\begin{figure*}
\centerline{\includegraphics[height=0.95\textwidth,angle=-90]{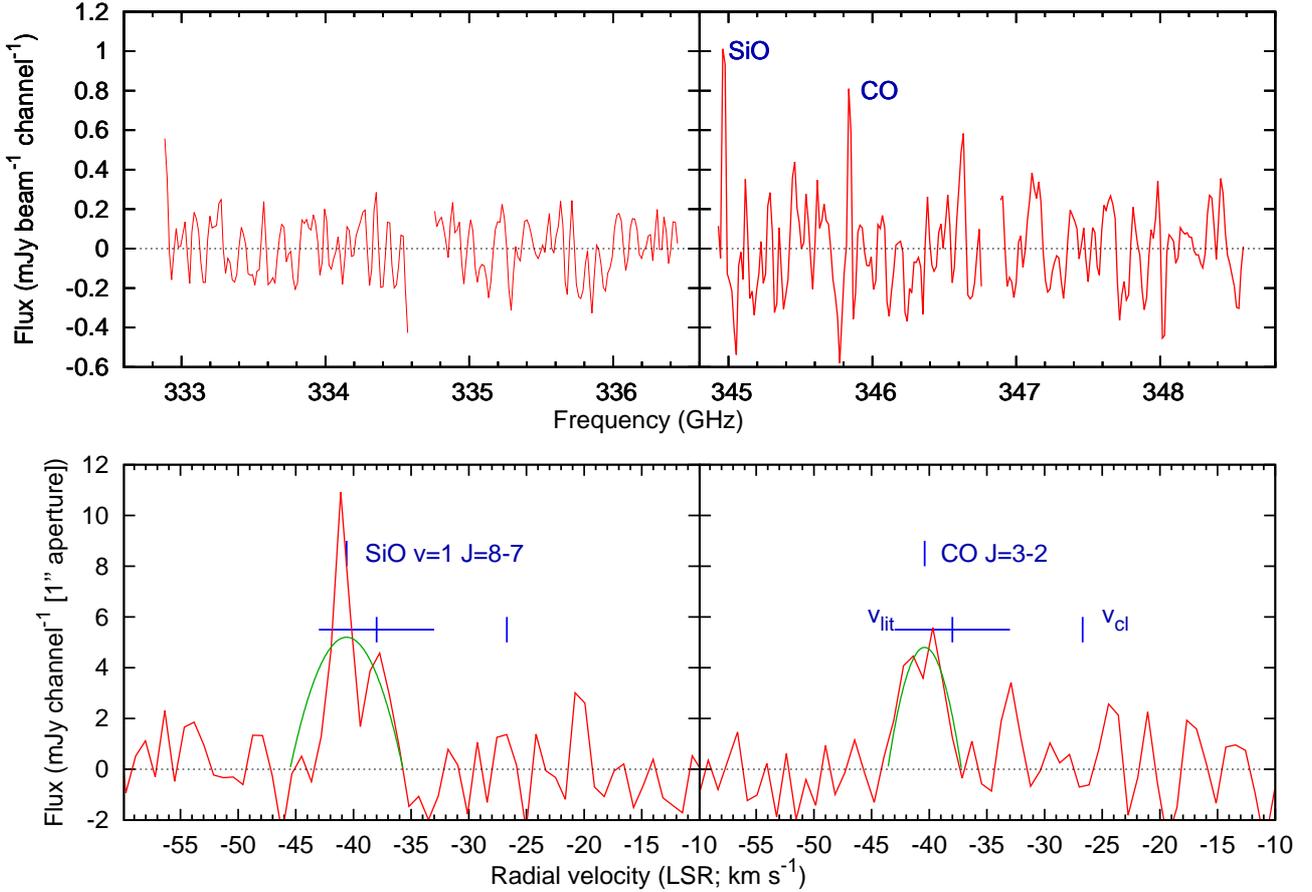}}
\caption{Line spectra from the observed position of V3. The spectral window containing the CO and SiO lines (marked) has been degraded in resolution to match the other channels. The lower panels show extracted regions around these lines at full resolution, extracted in a 1$^{\prime\prime}$ aperture around V3. A parabolic fit to the lines is also displayed, along with the mean optical velocity ($v_{\rm lit}$) and its range \citep{LW05} and the cluster velocity ($v_{\rm cl}$; \citealt{Harris96}).}
\label{SpecFig}
\end{figure*}

\begin{figure}
\centerline{\includegraphics[height=0.47\textwidth,angle=-90]{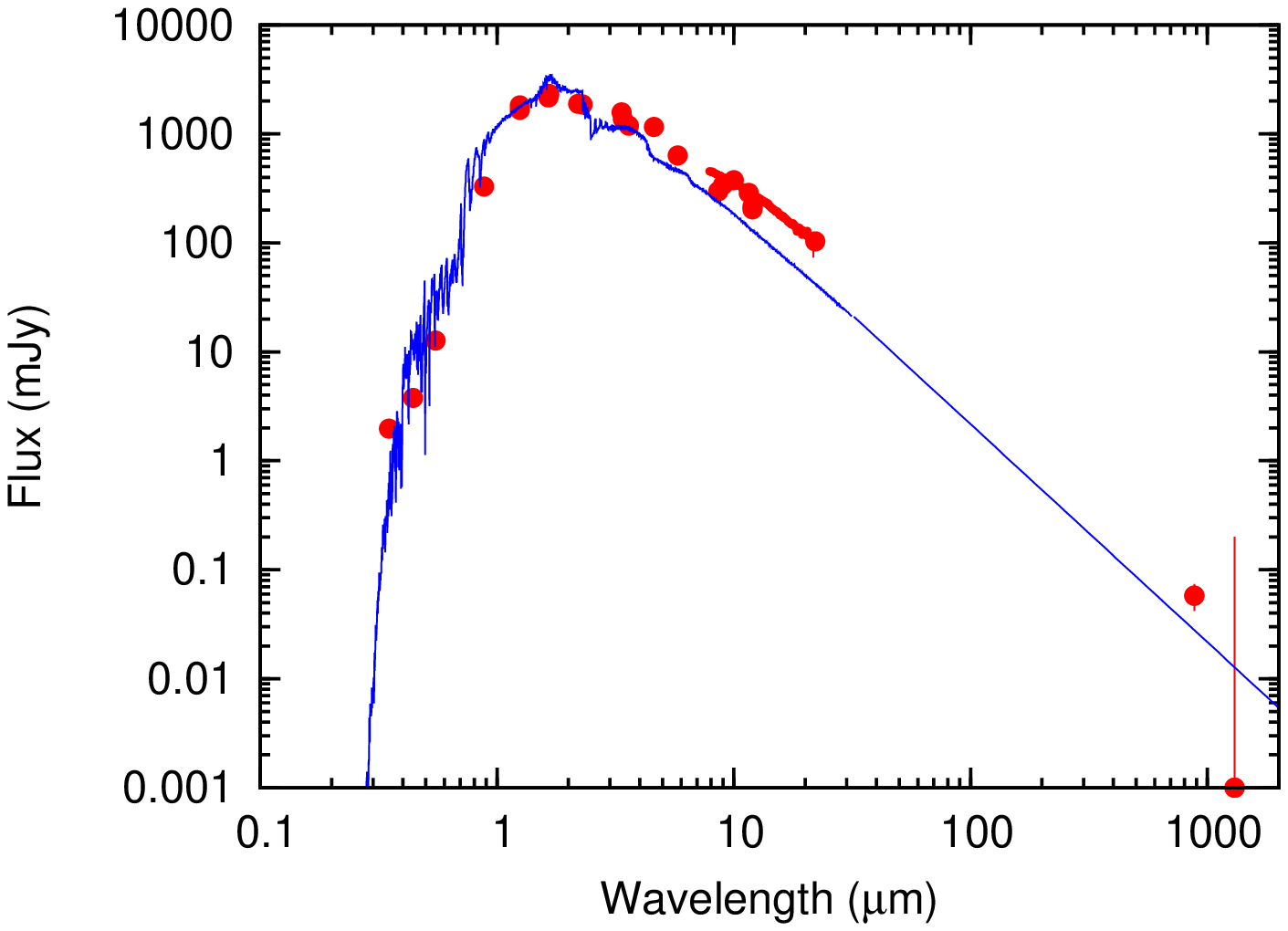}}
\caption{Spectral energy distribution of V3 (red points), and a dustless {\sc bt-settl} model atmosphere (blue line) at 3200 K, [Fe/H] = --1.0 dex and log($g$) = 0 dex \citep{AGL+03}. Photometry from the Magellanic Clouds Photometric Survey \citep{ZHT+02}; the Two-Micron All-Sky Survey 6X \citep[][VizieR catalogue II/281]{SCS+06}; \citet{GF73}; the \emph{Wide-field Infrared Survey Explorer} ``AllWISE'' catalogue \citet{CWC+13}; \citet{BvLM+10}; \citet{vLMO+06}; \citet{ITM+07}; \citet{MBvLZ11}; \citet{OFFPR02}; and \citet{MZL+15}.}
\label{SEDFig}
\end{figure}

ALMA (project code 2016.1.00078.S) observed 47 Tuc V3 in Band 7 on 2018 May 12, 13 and 15, totalling 5.2 hours of on-source observation over four spectral windows, spanning 331.74--348.73 GHz. The CO $J$=3--2 line (345.796 GHz) was covered at a resolution of 488 kHz (0.42 km s$^{-1}$), with a resulting noise of 0.62 mJy channel$^{-1}$ beam$^{-1}$, in a window that also covered the SiO $v = 1$ $J = 8-7$ (344.917 GHz) line. The remaining windows were observed at a resolution of 15.625 MHz ($\sim$13.5 km s$^{-1}$), attaining a total continuum sensitivity of 21 $\mu$Jy beam$^{-1}$. The $uv$-coverage of the array is sensitive to angular resolutions between 1.14 and 5.51$^{\prime\prime}$.

Figure \ref{SpecFig} presents the resulting spectra, extracted in a circle of radius 1$^{\prime\prime}$ around V3. Channel maps of the (stronger) SiO line show V3 is recovered as a point source at $00^h 25^m 16.^{\!s}023$ $-72^\circ 03^\prime 54\farcs70$, well within a synthesised beamwidth of its expected position ($00^h 25^m 16.^{\!s}003$ $-72^\circ 03^\prime 54\farcs68$; \citealt{GaiaDR2}). The CO line appears to be narrow, but perhaps with an additional redshifted wing, closer towards the cluster rest velocity, at a significance of 1.5 $\sigma$. The SiO line may be a maser source, and it may also lie on a weak plateau over a wider ($\sim\pm$4 km s$^{-1}$) extent. No lines appear in the other spectral windows.

The CO line peaks at 2.4 $\pm$ 0.6 mJy beam$^{-1}$ channel$^{-1}$ (3.8$\sigma$), with an integrated line strength of 35.8 $\pm$ 5.1 mJy km s$^{-1}$ (7$\sigma$) computed over the 1$^{\prime\prime}$ aperture and the velocity range --46 to --36 km s$^{-1}$. The half-width at zero power (FWZP) is $\sim$8 channels ($\sim$3.4 km s$^{-1}$). The same values for the SiO line are 8.2 $\pm$ 0.6 mJy beam$^{-1}$ channel$^{-1}$ (13$\sigma$), 48.9 $\pm$ 4.4 mJy km s$^{-1}$ (11$\sigma$), with a HWZP of 9 channels (3.8 km s$^{-1}$). The CO and SiO spectra over the 1$^{\prime\prime}$ aperture were each fit by a parabolic line function using $\chi^2$ minimisation, resulting in respective peak fluxes of 4.8 $\pm$ 0.6 and 5.2 $\pm$ 1.2 mJy, central velocities of --40.6 $\pm$ 0.3 and --40.4 $\pm$ 0.4 km s$^{-1}$, HWZP of 3.2 $\pm$ 0.4 and 4.9 $\pm$ 0.6 km s$^{-1}$, and reduced $\chi^2$ of 2.2 and 5.3.
% Offset 761.7 km/s

V3 is recovered in the frequency-averaged continuum image (the CO and SiO lines were masked from its construction), with a 345-GHz flux density of $F = 58 \pm 16$ $\mu$Jy ($\sim$3.5$\sigma$). Figure \ref{SEDFig} shows the spectral energy distribution of V3, accompanied by a dustless photospheric model for the underlying star. By scaling this model, the photospheric flux density at 345 GHz (869 $\mu$m) is expected to be $\sim$28 $\mu$Jy, other (unresolved) molecular lines can be expected to contribute $\sim$10 $\mu$Jy to the continuum flux density, and some synchrotron component may be expected from the chromosphere \citep{RM97}, meaning the continuum flux density from dust is $\lesssim$20 $\pm$ 16 $\mu$Jy. Unfortunately, this flux density does not greatly constrain the dust properties, except that its spectral emissivity slope between 22 and 869 $\mu$m is $\beta \gtrsim 0.02$ at 1$\sigma$. An unrelated 345-GHz point source is detected in the continuum at near the north-west edge of the synthesised image, at a distance of 10$\farcs$7, at $00^h 25^m 14.^{\!s}210$ $-72^\circ 03^\prime 48\farcs22$, with a flux density at 345 GHz of $F = 556 \pm 56$ $\mu$Jy. Another source is tentatively identified, closer to V3 ($2\farcs55$ west), at $00^h 25^m 15.^{\!s}479 -72^\circ 03^\prime 54\farcs60$, with flux density $F = 94 \pm 20$ $\mu$Jy. No counterparts of any kind exist within 2$^{\prime\prime}$ of those positions in the Centre de Donn\'ees astronomiques de Strasbourg database, except objects likely to be directly associated with V3 itself.

\section{Discussion \& Conclusions}

\begin{center}
\begin{table*}
\caption{Properties V3 and comparable Galactic objects}
\label{StarsTable}
\begin{tabular}{lcccccccccc}
    \hline \hline
    \multicolumn{1}{c}{Cluster} & \multicolumn{1}{c}{Type} & \multicolumn{1}{c}{$T_{\rm eff}$} & \multicolumn{1}{c}{$L$}         & \multicolumn{1}{c}{$P$} & \multicolumn{1}{c}{$\Delta V$} & \multicolumn{1}{c}{$K-[22]$} & \multicolumn{1}{c}{$\dot{M}$} & \multicolumn{1}{c}{$v_\infty$}    & \multicolumn{1}{c}{$I_{4.5(3\rightarrow2)}$}\\
    \multicolumn{1}{c}{\ }      & \multicolumn{1}{c}{\ }   & \multicolumn{1}{c}{(K)}           & \multicolumn{1}{c}{(L$_\odot$)} & \multicolumn{1}{c}{(d)} & \multicolumn{1}{c}{(mag)}      & \multicolumn{1}{c}{(mag)}    & \multicolumn{1}{c}{M$_\odot$ yr$^{-1}$} & \multicolumn{1}{c}{(km s$^{-1}$)} & \multicolumn{1}{c}{(Jy km s$^{-1}$)} \\
    \hline
V3    & M & 3153 & 2975 & 192 & 4.2 & $\sim$1.0 & $\sim$2 $\times$ 10$^{-7}$ & $\sim$3.2 & 0.036 \\
X Cnc & C & 2200 & 2800 & 195 & 1.9 &      1.03 &       7 $\times$ 10$^{-8}$ &        6.5 & 0.25 \\ % 2.2*29.4*(280/4500)^2
W Ori & C & 2600 & 3500 & 212 & 4.2 &      1.33 &       7 $\times$ 10$^{-8}$ &       10.  & 0.40 \\ % 4.1*41*(220/4500)^2
R Cas & M & 2800 & 3500 & 430 & 8.8 &      2.39 &       5 $\times$ 10$^{-7}$ &       10.  & 0.35 \\ % 7.8*29.7*(176/4500)^2
\hline
\end{tabular}
\end{table*}
\end{center}

These observations are ten times more sensitive than that previously obtained for CO $J$=2--1, allowing us to detect the CO $J$=3--2 line and SiO $v=1$ $J=8-7$ line. This can be compared to similar detections around Galactic (thin disk) stars \citep[e.g.][]{TKTY17,KDR+18}, to dust modelling of V3 itself \citep{MBvLZ11}, and in the context of pulsating giants in general \citep{DTJ+15,MZ16}. With this, we can infer several facts about metal-poor stellar winds, and about globular clusters.

\emph{Comparison to Galactic stars:} Identifying comparable Galactic stars is difficult, as no star has a similar combination of temperature, luminosity, and pulsation period and amplitude as V3. Table \ref{StarsTable} lists some of the closest comparisons which have been observed in CO $J$=3--2, along with their chemical types (C or M-type), temperatures, luminosities, periods, amplitudes, wind-expansion velocities, and CO $J$=3--2 line intensities (scaled to a distance of 4.5 kpc).

\emph{An SiO maser?} The 344 GHz $v=1$ line forms much closer to the star \citep{TKTY17}. It is normally weaker than the CO line, but the Si/C ratio in V3 is approximately twice that of Galactic stars \citep{RCGS14}. The sharp peak could indicate a near-stationary, molecular layer near the stellar surface. However, the SiO line is asymmetric and not well fit by a parabola (Figure \ref{SpecFig}), suggesting the line is more likely masing. Thermal wings may extend closer to v$_\infty$, but the true extent may be sensitivity-limited.

\emph{The ISRF:} \citet{MZL+15} constrained the mass-loss rate of V3 to be $\sim$1.2--3.5 $\times$ 10$^{-7}$ M$_\odot$ yr$^{-1}$. Galactic stars with similar mass-loss rates, luminosity and pulsation properties would (at 4.5 kpc distance) have CO $J$=3--2 flux densities of nearly 1 Jy km s$^{-1}$ (Table \ref{StarsTable}, data from \citet{SRO+13}; see also \citet{DBDdK+10,MdBZL18}). We retrieve a CO line intensity an order of magnitude lower than this. V3 is comparatively metal-poor, but the lower CO abundance should be approximately balanced by the higher flux expected from a slower wind ($I_{\rm CO} \appropto v_\infty^{-1.2}$; \citealt{DBDdK+10}), hence we anticipate that the CO line intensity is diminished due to strong photodissociation of CO \citep[cf.][]{MGH88}.

\citet{MZL+15} estimated that the ISRF at V3 was $\sim$9$\times$ stronger than in the Solar Neighbourhood, with a factor $\sim$3 uncertainty, matching this order-of-magnitude deficit in CO line flux. The same paper presented a model of CO dissociation around V1, which is expected to have an ISRF $\sim$5$\times$ stronger than V3. A CO $J$=3--2 intensity of 6.1 mJy km s$^{-1}$ was predicted, $\sim$5$\times$ smaller than that observed in V3. Given the myriad uncertainties in both the scaling and the modelling, it is remarkable that the CO line flux from V3 should so precisely match the expectations for a stellar wind experiencing dissociation by the ISRF. Based on this model, which assumed $v_\infty = 10$ km s$^{-1}$, half of the CO around V3 would be dissociated by $\sim$1000 R$_\ast$ ($\sim$0$\farcs$2), or after $\sim$2000 yr. It can be presumed that dust also becomes photo-ionised, leading to its absence in the ISM of most globular clusters \citep[e.g][]{BMvL+08}. While many mechanisms have been proposed to remove intra-cluster material from globular clusters \citep[see discussion in][]{MZ15a}, these observations can be interpreted as supporting this photo-evaporation model.

\emph{Dust and the wind-driving mechanism:} The available information does not allow us to directly determine whether radiation pressure on dust, momentum transfer from pulsations, or another mechanism (e.g., magnetic fields; \citealt[e.g.][]{DHA84}) physically drives the wind, hence we must look at the balance of evidence. Pulsation-driven winds have received little attention in the literature. However, the observations have similarities with the type B models of \citet{WlBJ+00}, which predict $\sim$10$^{-7}$ M$_\odot$ yr$^{-1}$ winds at $\sim$5 km s$^{-1}$ from stars where radiation pressure on dust does not exceed gravity until several stellar radii: pulsations would be needed to levitate material to this altitude. Magnetic fields cannot be completely ruled out either, however it would be conceptually difficult to generate the sudden factor of $\sim$100$\times$ increase in mass-loss rate seen during the transition to a dust-producing wind \citep{MdBZL18}. A wind driven by radiation pressure on dust can be examined in more detail.

As the CO line profile is well-represented by a paraboloid, $v_\infty$ can be taken directly as half the model's HWZP: 3.2 $\pm$ 0.4 km s$^{-1}$. However, if the tentative redshifted component is real, either $v_\infty$ could be slightly higher, or the CO envelope may be interacting with the lower-velocity interstellar media of the cluster.

The critical test of the role of dust in driving the wind is the metallicity scaling: if the wind momentum is derived from radiation pressure on dust, and in the absence of any other changes, the terminal wind velocity should scale as $v_\infty \appropto \sqrt{Z}$; \citep[e.g.][]{HTT94,NIE99}. If the wind momentum is set by a metallicity independent mechanism (e.g., pulsations or magnetic fields acting on gas), then there should be no such metallicity scaling, and the wind velocity will instead depend on the strength of this alternative driving mechanism. The observed wind velocity of V3 is a factor of 2--3$\times$ slower than the typical $v_\infty$ for Galactic (solar-metallicity) stars of similar properties (Table \ref{StarsTable}). As the metallicity of 47 Tuc is approximately 0.2 $Z_\odot$, this is closely in keeping with a $v_\infty \propto \sqrt{Z}$ law, meaning the wind is likely to be driven by radiation pressure on dust.

The extremely slow $v_\infty$ means that the wind of V3 may not become a net supersonic outflow until several tens of R$_\ast$. Driving a wind from this point is difficult: we expect the levitating effects of pulsation shocks to be minimal, for dust formation to be essentially complete, and for the wind to have reached its terminal velocity \citep[e.g.][]{LHNE16}. We therefore expect some infall of material back toward the star, even at large radii. Multiple passages through the condensation zone for a dust species may allow dust formation out to larger radii, allowing different species to condense, and resulting in V3's unusual, featureless dust spectrum. Additional species may include metallic iron \citep{MSZ+10}, which could provide the extra opacity needed to efficiently drive a wind via radiation pressure on dust.

The recovered $v_\infty$ exactly reproduces the 3.2 km s$^{-1}$ predicted for a wind accelerated purely by radiation pressure on metallic iron dust \citep{MBvLZ11}. The modelled mass-loss rate (9.4 $\times$ 10$^{-7}$ M$_\odot$ yr$^{-1}$) is higher than the constraint from evolutionary arguments ($\sim$1.2--3.5 $\times$ 10$^{-7}$ M$_\odot$ yr$^{-1}$). While this discrepancy could arise from a poor understanding of early-AGB mass-loss rates, a more obvious candidate is a more opaque form of dust, such as smaller or more porous dust grains, or a different type of condensate. Several of these factors may be needed to match the modelled mass-loss rate with the evolutionary one. We hypothesise that grains grow slowly in the metal-poor environment, before becoming large and opaque enough to be accelerated from the star. This gives a denser, quasi-stationary, dust-forming layer, with a higher opacity, making the mass-loss rate appear larger than it really is. This clumpy wind model is also similar to the 3D model of \citet{TKTY17}, where local acceleration causes denser, more-opaque clumps to achieve escape velocity once sufficient dust has formed, while the remainder cannot be efficiently ejected.

We caution that it is difficult to ascertain with certainty the wind-driving mechanism in metal-poor stars from observations of a single object, especially given the possibility of geometric asymmetries or time-dependencies in the wind properties. However, to summarise the results of these observations and those in the Magellanic Clouds, we find that low metallicity has no discernable impact on the predicted mass-loss rate of stars, nor the infrared excesses they obtain at a given evolutionary stage. This suggests that the mass-loss rate is set by stellar pulsations \citep[cf.][]{MZ16,MdBZL18}. However, the low outflow velocity from V3 compared to Galactic stars (also suggested in the Magellanic Clouds; \citet{GVM+16,MSS+16,GvLZ+17}) and change in dust properties suggests that $v_\infty$ is still set by a metallicity dependent mechanism, namely radiation pressure on dust. Observation at higher frequencies, to detect higher CO transitions and the dust continuum, would help constrain both the dust properties (e.g., the spectral slope of the dust emission) and the ISRF; observations of stars at lower metallicity would help constrain the metallicity scaling law; while observations of Halo stars are encouraged to explore the effects of the ISRF.

\section*{Acknowledgements}

We thank Livia Origlia for her constructive critical review of this Letter. IM and AAZ acknowledge support from the UK Science and Technology Facility Council under grants ST/L000768/1 and ST/P000649/1. This paper makes use of the following ALMA data: ADS/JAO.ALMA\#2016.1.00078.S. ALMA is a partnership of ESO (representing its member states), NSF (USA) and NINS (Japan), together with NRC (Canada), MOST and ASIAA (Taiwan), and KASI (Republic of Korea), in cooperation with the Republic of Chile. The Joint ALMA Observatory is operated by ESO, AUI/NRAO and NAOJ.


%\bibliographystyle{mnras}
%\bibliography{references} % if your bibtex file is called example.bib

%%%%%%%%%%%%%%%%% APPENDICES %%%%%%%%%%%%%%%%%%%%%

%\appendix

% Don't change these lines
\bsp	% typesetting comment
\label{lastpage}
\end{document}